\documentclass[article,singlecolumn]{elsarticle}

\journal{Materials Chemistry and Physics}

\usepackage{lineno,hyperref}
\usepackage[version=3]{mhchem} 
\usepackage[T1]{fontenc}       
\usepackage[version=3]{mhchem} 
\usepackage[T1]{fontenc}       
\usepackage{graphicx}
\usepackage{dcolumn}
\usepackage{bm}
\usepackage{xcolor}
\usepackage{amsmath}
\usepackage{setspace}
\usepackage{chemformula}
\usepackage{soul}
\newcommand{\C}{$^\circ$C}
\newcommand{\TC}{$T_{\text C}$}
\newcommand{\TSR}{$T_{\text SR}$}
\newcommand{\TN}{$T_{\text N}$}
\newcommand{\JKK}{JKg$^{-1}$K$^{-1}$}
\newcommand{\JK}{JKg$^{-1}$}

\newcommand{\Hmax}{$H_{\text{max}}$}

\begin{document}

\begin{frontmatter}

\title{Characterization of the Magnetocaloric Effect in \ch{$R$Mn6Sn6} including High-Entropy Alloys}
\author[mainaddress]{Kyle~Fruhling\corref{mycorrespondingauthor}}\cortext[mycorrespondingauthor]{Corresponding author}
\ead{fruhling@bc.edu}
\author[mainaddress]{Xiaohan~Yao}
\author[mainaddress]{Alenna~Streeter}
\author[mainaddress]{Fazel~Tafti}
\address[mainaddress]{Department of Physics, Boston College, Chestnut Hill, MA 02467, USA}

\begin{abstract}
We present a comprehensive study of the magnetocaloric effect (MCE) in a family of kagome magnets with formula \ch{$R$Mn6Sn6} ($R$=Tb, Ho, Er, and Lu).
These materials have a small rare-earth content and tunable magnetic ordering, hence they provide a venue to study the fundamentals of the MCE.
We examine the effect of different types of order (ferrimagnetic and antiferromagnetic) and the presence of a metamagnetic transition on the MCE.  
We extend the study to high-entropy rare-earth alloys of the family. Our results suggest several guidelines for enhancing the MCE in tunable magnetic materials with a small rare-earth content.
\end{abstract}
\begin{keyword}
Magnetocaloric effect\sep Kagome lattice\sep Rare-earth
\end{keyword}
\end{frontmatter}


\section{Introduction}

Since the discovery of a giant magnetocaloric effect (MCE) in \ch{Gd5Si2Ge2} by Pecharsky and Gschneidner~\cite{pecharsky_giant_1997}, magnetic refrigeration has been considered as a viable replacement for conventional gas refrigeration.
Engineering designs already exist for magnetic refrigerators that are at least 20\% more energy efficient than gas refrigerators~\cite{yu_review_2010,alahmer_magnetic_2021}.
The higher energy efficiency as well as elimination of greenhouse and ozone-depleting gases make magnetic refrigeration a critical aspect of sustainability.
However, there are several challenges related to magnetocaloric materials (MCMs) that must be addressed before magnetic refrigerators can be commercialized. 
(i) Current benchmark MCMs are either elemental Gd and Ho or rare-earth-rich compounds such as \ch{Gd5Si2Ge2}, and \ch{HoB2}~\cite{terada_high-efficiency_2021,luo_exploring_2018,gschneidner_magnetocaloric_2000,franco_magnetocaloric_2012}. 
Since using large amounts of rare-earth is not sustainable, several attempts have been made to find rare-earth-free MCMs such as AlFe$_2$B$_2$ and MnAs~\cite{tan_magnetocaloric_2013,wada_giant_2003,tegus_transition-metal-based_2002}.
Unfortunately, these materials are either less efficient or contain toxic elements such as As and P.
(ii) The giant MCE in rare-earth-rich materials originates from a large entropy release in a narrow temperature range across a first-order magneto-structural phase transition (FOPT). 
The hysteresis in a FOPT limits thermal cycling of the material. 
(iii) Recently, disorder has been used to broaden the magnetic transition and enhance the refrigerant capacity in Heusler compounds and high-entropy alloys~\cite{jones_effect_2012,ucar_effect_2014,amaral_disorder_2014,pal_enhancing_2020}.
However, disorder works against structural integrity of MCMs and is a challenging parameter to control during manufacturing processes.
(iv) Few microscopic mechanisms besides FOPT and disorder have been proposed to improve the MCE. 
A poor theoretical understanding of the microscopic mechanisms that control the MCE severely limits the pace of discovering high-efficiency MCMs.

\begin{figure*}
  \includegraphics[width=\textwidth]{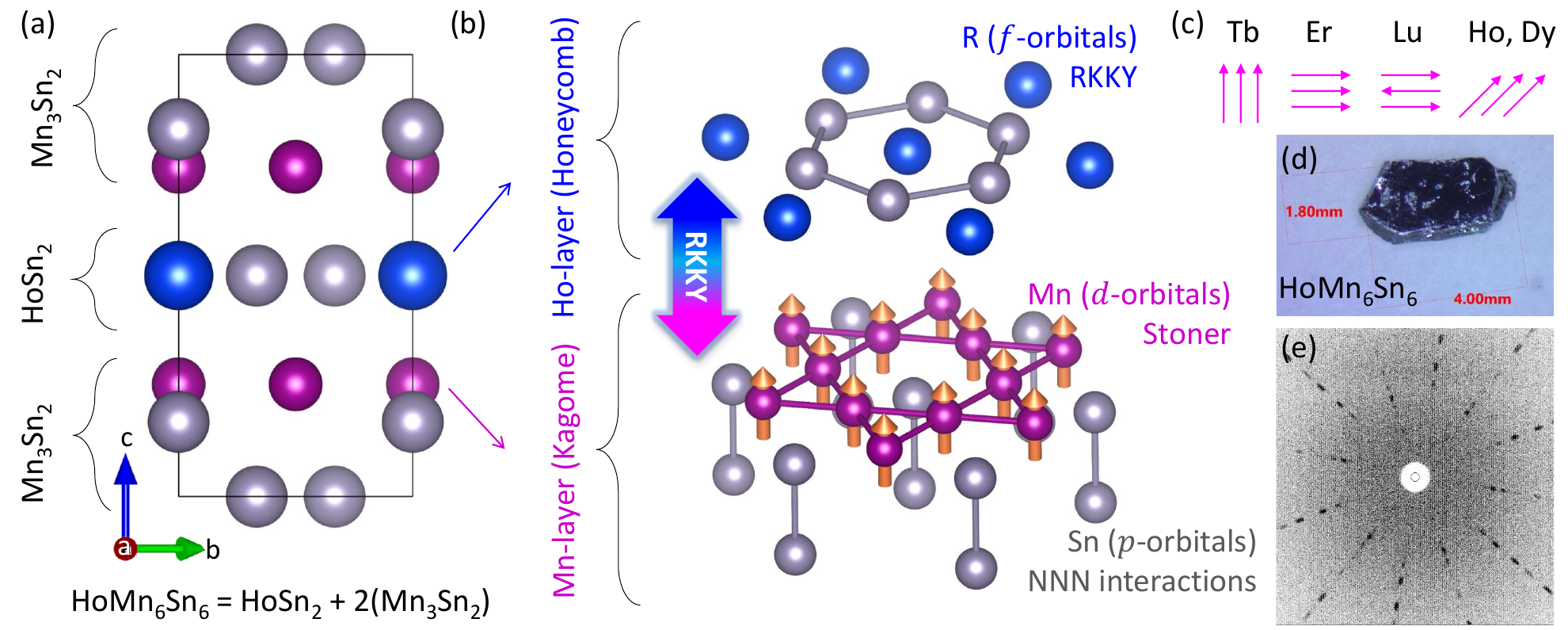}
  \caption{\label{fig:CIF}
  (a) Each unit cell of \ch{HoMn6Sn6} (a representative member of \ch{$R$Mn6Sn6} family) contains one honeycomb layer of \ch{HoSn2} sandwiched between two kagome layers of \ch{Mn3Sn2}.
  (b) The electronic structure consists of itinerant $p$ and $d$ electrons as well as localized $f$ electrons provided by Sn, Mn, and Ho, respectively.
  (c) Magnetism in the Mn layers varies significantly by changing the rare-earth due to RKKY interactions.
  (d) Optical image of a large single crystal.
  (e) Laue diffraction along the crystallographic $c$-axis confirms the hexagonal unit cell ($P6/mmm$) and absence of twinning in the crystal.
  }
\end{figure*}

In this work, we use a tunable family of \ch{$R$Mn6Sn6} materials ($R$=Tb, Ho, Er, and Lu) to gain a deeper insight into the fundamentals of the MCE.
Each unit cell of \ch{$R$Mn6Sn6} comprises one layer of \ch{$R$Sn_2} sandwiched between two layers of \ch{Mn3Sn2} (Figs.~\ref{fig:CIF}a,b).
The manganese atoms form a kagome lattice and rare-earth atoms form a triangular lattice.
Only 1 out of 13 atoms in the chemical formula is a lanthanide.
The small rare-earth content ($<8\%$) and absence of toxic elements in the composition make these materials a sustainable platform for the MCE. 
They host a variety of magnetic structures (Fig.~\ref{fig:CIF}c) due to Ruderman-Kittel-Kasuya-Yosida (RKKY) interactions between the itinerant electrons (Mn-$d$ and Sn-$p$ electrons) and local moments (rare-earth $f$ electrons)~\cite{yin_quantum-limit_2020,li_discovery_2023}.
Magnetism can be easily tuned from ferrimagnetic (FIM) to antiferromagnetic (AFM) by changing the rare-earth element as illustrated in Fig.~\ref{fig:CIF}c.
We study the effect of different magnetic structures on the MCE by comparing the magnetic entropy change ($\Delta S_m$), full width at half maximum (FWHM) of $-\Delta S_m$, and refrigerant capacity ($RC$) across the high-temperature magnetic transition of \ch{TbMn6Sn6}, \ch{HoMn6Sn6}, \ch{ErMn6Sn6}, \ch{LuMn6Sn6}, and their high-entropy alloys.
These compounds also have one or two low-temperature transitions.
We present the low-temperature MCE as supplemental material since it is not relevant for technological applications.

\section{Experimental methods}
\subsection{Material synthesis}
Single crystals of \ch{$R$Mn6Sn6} were grown using a self-flux technique.
The starting chemicals were terbium, holmium, erbium, and lutetium pieces (99.9\%), manganese granules (99.98\%), and tin pieces (99.999\%).
They were mixed with the ratio $R$:Mn:Sn = $X$:6:18 ($X$ = 1.25 for Tb, 1.50 for Ho, 1.10 for Er, and 1.20 for Lu), placed in alumina crucibles and sealed inside evacuated silica tubes.
The tubes were heated in a box furnace to 1000\C\ at 3\C/min, held at that temperature for 12h, cooled to 600\C\ at 6\C/h, and centrifuged to remove the excess flux.
Using lesser amounts of rare-earth or centrifuging at $T<600$\C\ resulted in the formation of impurity phases such as HoSn$_2$ and $R$MnSn$_2$.

The crystallographic structure was verified by Laue x-ray diffraction using a real-time back-reflection MWL120 system. 
Picture of a representative \ch{HoMn6Sn6} crystal and its Laue diffraction pattern are shown in Figs.~\ref{fig:CIF}d,e.
The chemical composition of each material was confirmed by energy dispersive x-ray spectroscopy (EDX) in an FEI Scios DualBeam electron microscope equipped with Oxford detector.
The EDX spectra and composition analyses of \ch{$R$Mn6Sn6} samples are presented in the supplemental Fig.~S1 and Table.~S1. 

\subsection{Magnetic measurements}
Magnetization measurements were performed on single crystals using a 7~T Quantum Design MPMS-3 equipped with vibrating sample magnetometry (VSM) and oven options.
The MCE was quantified by measuring magnetization isotherms in the vicinity of the critical temperature and evaluating the magnetic entropy change ($\Delta S_m$) using the Maxwell equation~\cite{kinami_magnetocaloric_2018,abramchuk_tuning_2019},
\begin{equation}
\label{eq:maxwell}
\Delta S_m (T, \Delta H) = \int_0^{H_{\text{max}}} \left(\frac{\partial M}{\partial T} \right)_H dH
\end{equation}
which was approximated with the following expression for discrete isotherms:
\begin{equation}
\label{eq:numerical}
\Delta S_m \left(T=\frac{T_1+T_2}{2}\right) = \left(\frac{1}{T_2-T_1}\right)\times\left[ \int_0^{H_{\text{max}}} M(T_2,H)\, dH - \int_0^{H_{\text{max}}} M(T_1,H)\, dH \right]
\end{equation}
using $H_{\text{max}}=$ 0.5, 2, and 5~T.
A bell-shaped curve was obtained for $-\Delta S_m(T)$ with three parameters that characterize the MCE: (i) the height of the curve ($-\Delta S_\text{MAX}$), (ii) full width at half maximum (FWHM), and (iii) refrigerant capacity ($RC$) defined as the area under the curve.
\begin{equation}
\label{eq:rc}
RC = \int_{\text{FWHM}} -\Delta S_m(T)\, dT 
\end{equation}

\section{Results and Discussion}
\subsection{Temperature Dependence of Magnetic Susceptibility}
\begin{figure}
\centering
  \includegraphics[width=0.8\textwidth]{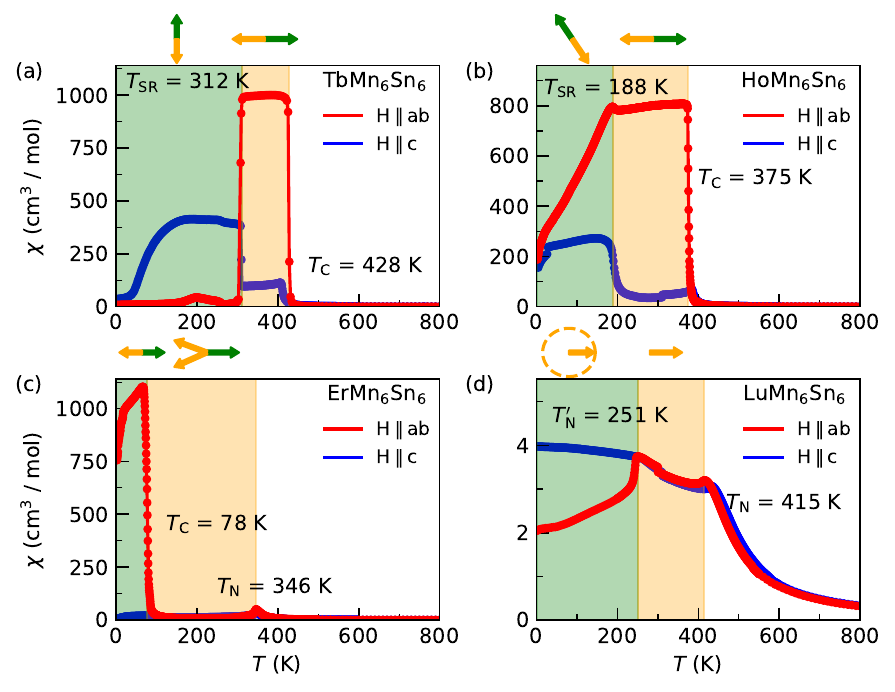}
  \caption{\label{fig:SUSC}
    Temperature dependence of the zero-field cooled magnetic susceptibility in (a) \ch{TbMn6Sn6}, (b) \ch{HoMn6Sn6}, and (c) \ch{ErMn6Sn6} measured at $\mu_0H=100$~Oe, and (d) \ch{LuMn6Sn6} measured at $\mu_0H=5000$~Oe.The green arrows represent the direction of the rare-earth moment and the orange arrows represent the direction of the Mn moment.
  }
\end{figure}
The magnetic ordering of \ch{$R$Mn6Sn6} depends sensitively on the choice of rare-earth element, evidence of a strong $R$-Mn coupling due to RKKY interactions.
Figure~\ref{fig:SUSC} shows temperature dependence of the magnetic susceptibility $\chi(T)$ in \ch{$R$Mn6Sn6} for $R$=Tb, Ho, Er, and Lu with $H\|ab$ and with $H\|c$.
All four compounds undergo different magnetic transitions at high ($T>300$K) and low ($T<300$K) temperatures.
In each case, the rare-earth and Mn moments, represented by the green and orange arrows respectively, are antiparallel.
In Fig.~\ref{fig:SUSC}a, \ch{TbMn6Sn6} shows an FIM transition with in-plane moments at the Curie temperature \TC=428~K.
It then undergoes a spin reorientation at \TSR=312~K where the moments rotate from in-plane to out-of-plane direction as confirmed by the susceptibility measurements with $H\|c$. 
In Fig.~\ref{fig:SUSC}b, \ch{HoMn6Sn6} shows an in-plane FIM transition at \TC=375~K and a spin reorientation transition at \TSR=188~K where the moments cant towards the c-axis~\cite{el_idrissi_magnetic_1991}.
In Fig.~\ref{fig:SUSC}c, \ch{ErMn6Sn6} shows an AFM transition at \TN=346~K and an FIM transition at \TC=78~, both with moments aligned in the $ab$-plane. 
It is shown recently~\cite{riberolles_low-temperature_2022} that the AFM order in \ch{ErMn6Sn6} has a spiral character, so that the Mn moments in adjacent layers have an angle relative to each other, and their net moment is canceled by the rare-earth moment, as represented by the arrows at the top of Fig.~\ref{fig:SUSC}c.
In Fig.~\ref{fig:SUSC}d, \ch{LuMn6Sn6} orders as a collinear in-plane anti-ferromagnet at \TN=415~K and transitions into an AFM flat-spiral structure of ferromagnetic sheets at \TN$'$=251~K~\cite{matsuo_study_2006}. 
Note that Lu$^{3+}$ is non-magnetic, hence the absence of a green vector in the caption of Fig.~\ref{fig:SUSC}d.
Our susceptibility data are consistent with previous reports on these materials~\cite{venturini_magnetic_1991,clatterbuck_magnetic_1999}.

\subsection{Field Dependence of Magnetization}
\begin{figure}
\centering
  \includegraphics[width=0.8\textwidth]{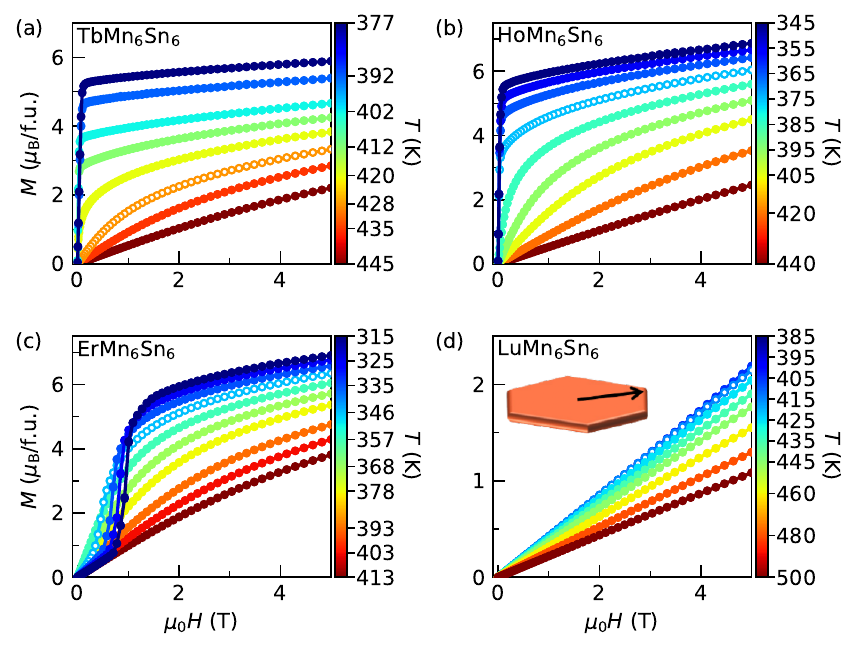}
  \caption{\label{fig:MH}
    Field dependence of the magnetization in (a) \ch{TbMn6Sn6}, (b) \ch{HoMn6Sn6}, (c) \ch{ErMn6Sn6}, and (d) \ch{LuMn6Sn6} measured at several temperatures indicated by color bars.
    All measurements were done with field in the $ab$-plane as illustrated in the inset of panel d.
    The temperatures quoted in color bars correspond to the isotherms in each panel.
    For each compound, the curve with empty symbols is taken at its magnetic transition temperature.
  }
\end{figure}
The field dependence of magnetization is summarized in Fig.~\ref{fig:MH} for \ch{$R$Mn6Sn6} with $R$=Tb, Ho, Er, and Lu. 
In Figs.~\ref{fig:MH}a and \ref{fig:MH}b, \ch{TbMn6Sn6} and \ch{HoMn6Sn6} show typical FIM magnetization curves that begin to saturate at approximately 0.2~T when $T\ll T_{\text C}$.

Unlike \ch{TbMn6Sn6} and \ch{HoMn6Sn6}, the magnetization of \ch{ErMn6Sn6} begins to saturate at nearly 1~T, instead of 0.2~T, following a metamagnetic (MM) transition.
The MM transition signals a change of magnetic state from AFM to FIM induced by the external magnetic field.
Below 1~T, the dark blue $M(H)$  curve in Fig.~\ref{fig:SUSC}c shows a linear field dependence at $T<T_{\text N}$ in \ch{ErMn6Sn6}, characteristic of the AFM state.
Then at 1~T, a spin-flop transition occurs and magnetization jumps to a higher value. 
Thus, the $M(H)$ behavior observed in Fig.~\ref{fig:MH}c implies in-plane moments and AFM ordering between the planes at $H=0$ followed by a spin-flop (MM transition) at around 1~T in \ch{ErMn6Sn6}.

In Fig.~\ref{fig:MH}d, \ch{LuMn6Sn6} shows linear $M(H)$ curves from 0 to 5~T, indicating an AFM ordering.
Similar to \ch{LuMn6Sn6}, other $R$\ch{Mn6Sn6} compounds with a non-magnetic rare-earth ($R$=Y, Sc) show a spiral AFM order~\cite{venturini_incommensurate_1996,ghimire_competing_2020,zhang_magnetic_2022}.

\subsection{Magnetocaloric Effect}
\begin{figure}
\centering
  \includegraphics[width=0.8\textwidth]{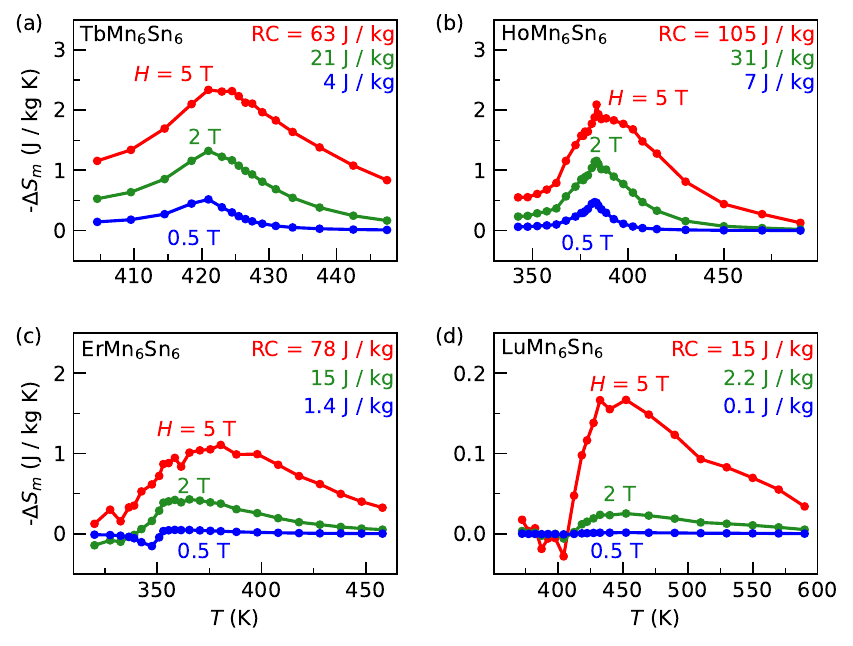}
  \caption{\label{fig:MCE}
   Magnetocaloric effect evaluated in (a) \ch{TbMn6Sn6}, (b) \ch{HoMn6Sn6}, (c) \ch{ErMn6Sn6}, and (d) \ch{LuMn6Sn6}. 
   The red, green, and blue curves correspond to the magnetic entropy integrated with \Hmax=5, 2, and 0.5~T, respectively, according to Eq.~\ref{eq:numerical}.
   All measurements are done with the magnetic field in the $ab$-plane.
  }
\end{figure}
We characterized the MCE by evaluating $\Delta S_m(T)$ from the $M(H)$ curves using Eq.~\ref{eq:numerical}.
The red, green, and blue curves in Fig.~\ref{fig:MCE} trace the change of entropy ($-\Delta S_m(T)$) across the magnetic phase transition under \Hmax=5, 2, and 0.5~T, respectively.
The bell-shaped curve for each compound peaks at a temperature near \TC\ or \TN, where the separation between $M(H)$ curves in Fig.~\ref{fig:MH} is maximum.
It is evident from Eq.~\ref{eq:numerical} that a larger separation between consecutive $M(H)$ isotherms means a larger entropy change.
For example, \ch{TbMn6Sn6} and \ch{HoMn6Sn6} show similar $-\Delta S_{\text{MAX}}$ values in Figs.~\ref{fig:MCE}a,b due to a comparable separation between their magnetization isotherms in Figs.~\ref{fig:MH}a,b.

Despite having comparable $-\Delta S_{\text{MAX}}$ values, \ch{HoMn6Sn6} is a more efficient MCM than \ch{TbMn6Sn6} because its bell-shaped $-\Delta S_m(T)$ curve has a larger FWHM. 
Thus, the refrigerant capacity ($RC$) defined as the area under the bell-shaped curve within FWHM (Eq.~\ref{eq:rc}) is greater in \ch{HoMn6Sn6} than in \ch{TbMn6Sn6}.
The $RC$ values are quoted for each compound in the upper right corner of every panel in Fig.~\ref{fig:MCE}.
Regardless of \Hmax, the $RC$ values for \ch{HoMn6Sn6} are larger than those for \ch{TbMn6Sn6} due to a larger FWHM. 

The significance of FWHM in the MCE is further illustrated in a comparison between \ch{TbMn6Sn6} and \ch{ErMn6Sn6} (Figs.~\ref{fig:MCE}a,c).
Although $-\Delta S_{\text {MAX}}$ in \ch{ErMn6Sn6} is nearly half the value in \ch{TbMn6Sn6} (1.1 vs. 2.3 \JKK), a much larger FWHM (wider bell-shaped curve) produces a larger $RC$ in in \ch{ErMn6Sn6} than in \ch{TbMn6Sn6} at \Hmax=5~T (78 vs. 63 \JK).

Despite having a larger $RC$ at \Hmax=5~T, \ch{ErMn6Sn6} has a smaller $RC$ than \ch{TbMn6Sn6} at \Hmax=2 and 0.5~T.
This is due to the AFM ordering in \ch{ErMn6Sn6} at low fields, which unlike the FIM state in \ch{TbMn6Sn6}, produces a small net magnetization.
As shown in Fig.~\ref{fig:MH}c, \ch{ErMn6Sn6} has a linear $M(H)$ curve from zero to 1~T before undergoing the MM transition that changes the shape of its $M(H)$ curves from linear to S-shaped (typical of FIM materials).
This field-induced transition from a low-field AFM to high-field FIM state is responsible for the change of the MCE in \ch{ErMn6Sn6} from small at \Hmax=0.5~T to large at \Hmax=2 and 5~T. 

In Fig.~\ref{fig:MCE}d, \ch{LuMn6Sn6} shows a very small $RC$ even at \Hmax=5~T because it remains in the AFM state from small to large fields.
The $M(H)$ curves of \ch{LuMn6Sn6} in Fig.~\ref{fig:MH}d remain linear with minimal separation between consecutive isotherms and without any MM transitions, hence the small MCE at all fields.

\subsection{High-Entropy Alloys}
\begin{figure}
\centering
  \includegraphics[width=0.8\textwidth]{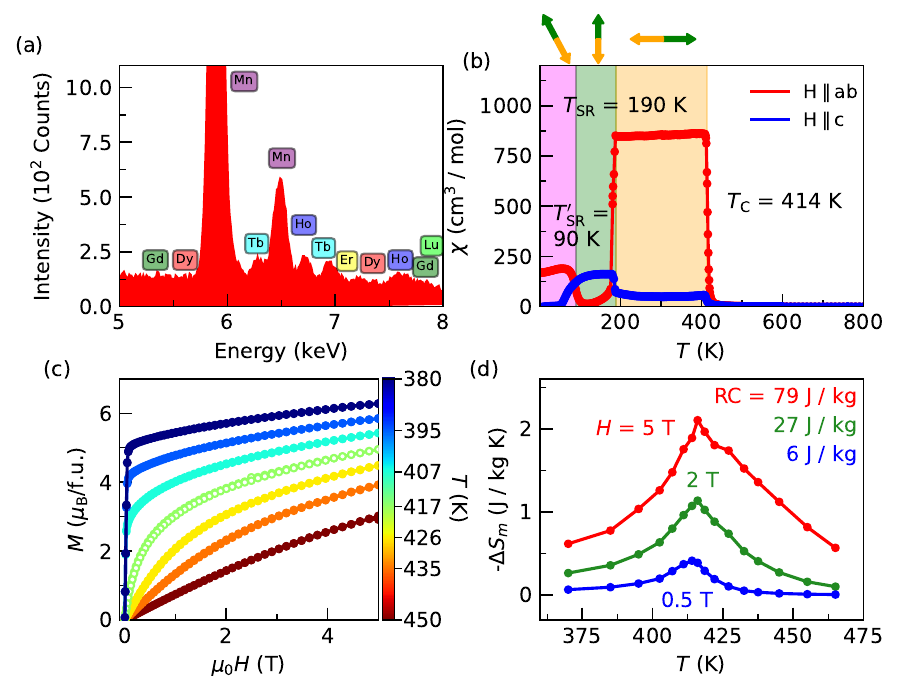}
  \caption{\label{fig:HEA-6}
   (a) EDX spectrum confirms the presence of six rare-earth elements in the high-entropy alloy (6R-HEA). 
   (b) Susceptibility versus temperature shows an FIM transition at high temperatures and two spin reorientation transitions at low temperatures. The green and orange arrows represent the rare-earth and Mn moments, respectively.
   (c) The Magnetization curves of 6R-HEA. 
   (d) The MCE in the 6R-HEA.
  }
\end{figure}
Following the work of Yeh \emph{et al.} in the early 2000s~\cite{yeh_nanostructured_2004}, high-entropy alloys (HEAs) have been used in applications that require extreme strength, oxidation resistance, and high temperature stability~\cite{george_high-entropy_2019}.
Recently, HEAs of transition-metals, such as MnFeCoNiCu and (FeCoNi)$_{60}$Cu$_{19}$Mn$_{21}$, have been tested for their magnetocaloric performance~\cite{perrin_mossbauer_2019,perrin_role_2017,kurniawan_curie_2016}.
The working hypothesis is that the high-entropy alloying procedure randomizes the exchange interactions and broadens magnetic transitions.
Therefore, it enhances $RC$ due to a larger FWHM according to Eq.~\ref{eq:rc}~\cite{perrin_mossbauer_2019,perrin_role_2017,kurniawan_curie_2016}.

In our study, we decided to examine the effect of high-entropy alloying on the MCE of \ch{$R$Mn6Sn6} system by mixing different rare-earth elements on the $R$ site~\cite{min_topological_2022}.
The EDX map in Fig.~\ref{fig:HEA-6}a confirms successful crystal growth of the HEA (Gd$_{0.21}$Tb$_{0.17}$Dy$_{0.15}$Ho$_{0.16}$Er$_{0.19}$Lu$_{0.13}$)Mn$_6$Sn$_6$, (see also the supplemental Table~S1) which we refer to as 6R-HEA. 
The susceptibility data in Fig.~\ref{fig:HEA-6}b shows an FIM transition at \TC=414~K followed by spin reorientation transitions at \TSR=190~K and \TSR$'$=90~K, respectively.
Using $M(H)$ curves near \TC\ in Fig.~\ref{fig:HEA-6}c, we computed the magnetic entropy change $\Delta S_m$ and plotted $-\Delta S_m$ as a function of temperature in Fig.~\ref{fig:HEA-6}d.
By integrating the area under the bell-shaped curves, we evaluated $RC$ at \Hmax=5, 2, and 0.5~T.
The $RC$ values reported in the inset of Fig.~\ref{fig:HEA-6}d are comparable to those of \ch{TbMn6Sn6} and less than those of \ch{HoMn6Sn6}.

The important observation in Fig.~\ref{fig:HEA-6} is that MCE in the HEA is not larger than that of all pure \ch{$R$Mn6Sn6} (compare Fig.~\ref{fig:HEA-6}d and Fig.~\ref{fig:MCE}).
This goes against the hypothesis that high-entropy alloying randomizes RKKY exchange interactions and increases $RC$ by broadening the magnetic transition.
Similar behavior is seen in alloys made with fewer rare-earth elements which we discuss in the supplemental material.

\section{Conclusions}
Comparing Fig.~\ref{fig:HEA-6} with \ref{fig:SUSC} and \ref{fig:MH} shows the HEA has a behavior similar to \ch{TbMn6Sn6} with an FIM ordering.
This FIM tendency could be expected given the FIM ordering observed in most of the rare-earth elements chosen.
Specifically, \ch{GdMn6Sn6}, \ch{TbMn6Sn6}, and \ch{DyMn6Sn6}~\cite{gorbunov_magnetic_2012,el_idrissi_magnetic_1991,chen_magnetic_2021} show in-plane FIM ordering while \ch{HoMn6Sn6} shows a canted FIM order~\cite{el_idrissi_magnetic_1991,malaman_magnetic_1999}. 
\ch{ErMn6Sn6} has an AFM order at high temperatures but that also changes into FIM order at lower temperatures~\cite{clatterbuck_magnetic_1999} or under field. 
Only \ch{LuMn6Sn6} shows AFM ordering in the region of interest.
Thus, HEAs made of these rare-earth elements have a tendency toward FIM ordering.
Additionally, the high-temperature magnetic transition is predominantly due to correlations within the Mn kagome lattice, unlike the low-temperature order which is controlled by the rare-earth sublattice~\cite{lee_interplay_2023}.
This can be seen in Figs.~\ref{fig:SUSC}a and \ref{fig:HEA-6}b, where the high-temperature \TC\ (controlled by Mn sublattice) is nearly the same between \ch{TbMn6Sn6} and the 6R-HEA, whereas the low-temperature \TSR\ (controlled by the rare-earth) is dramatically reduced from 312~K in \ch{TbMn6Sn6} to 190~K in 6R-HEA.

\begin{table}
 \caption{\label{tab:METRICS}Comparison of MCE metrics with \Hmax=5~T in \ch{$R$Mn6Sn6} materials investigated in this study. $T_c$ is the higher critical temperature (which could be either FIM or AFM).}
 \begin{tabular}[width=3.6in]{l|l|l|l|l}
 \hline
 \hline
 Material & T$_{c}$ (K) & -$\Delta S_{\text {MAX}}$(\JKK) & FWHM (K) & $RC$(\JK)                  \\
 \hline
 TbMn$_{6}$Sn$_{6}$  & 428 & 2.3  & 36  & 63  \\
 HoMn$_{6}$Sn$_{6}$  & 375 & 2.1  & 57  & 105  \\
 ErMn$_{6}$Sn$_{6}$  & 346 & 1.1  & 89  & 78  \\
 LuMn$_{6}$Sn$_{6}$  & 415 & 0.17 & 112 & 15  \\ 
 6R-HEA              & 414 & 2.1  & 52  & 79  \\
 \hline
 \hline
 \end{tabular}
 \end{table}

By comparing the MCE metrics in the different studied \ch{$R$Mn6Sn6} members in Table~\ref{tab:METRICS}, we arrive at several practical conclusions.
(i) FM and FIM transitions (e.g. in \ch{HoMn6Sn6}) are more efficient than the AFM transition (e.g. in \ch{LuMn6Sn6}) for generating a large MCE.
(ii) The maximum entropy release ($\Delta S_\text{MAX}$) is not the only decisive factor in the efficiency of a MCM. 
A wide transition, i.e. a large FWHM, is equally important in generating a large $RC$. 
(iii) High-entropy alloying could be a good mechanism to increase FWHM and $RC$, however, it is important that the alloying procedure truly randomizes the magnetic interactions. As we found in the case of \ch{$R$Mn6Sn6} systems, alloying on the rare-earth site did not randomize the RKKY interactions, because nearly all the rare-earth atoms in the $R$-site have a tendency toward FIM ordering.
The above points provide practical guidelines for materials scientists in their search for the next generation of magnetocaloric materials.

\section*{Acknowledgments}
The authors acknowledge funding by the National Science Foundation under Award No. DMR-2203512.
F.T. and X.Y. acknowledge funding by the Department of Energy, Office of Basic Energy Sciences, Division of Physical Behavior of Materials under award number DE-SC0023124 to perform EDX measurements. 
The authors thank Z.-C. Wang for the experimental assistance during early stages of the project.

\bibliography{Fruhling_Bibliography}

\end{document}


\begin{frontmatter}

\title{Supplemental Material:\\ Characterization of the Magnetocaloric Effect in \ch{$R$Mn6Sn6} including High-Entropy Alloys}
\author[mainaddress]{Kyle~Fruhling\corref{mycorrespondingauthor}}
\cortext[mycorrespondingauthor]{Corresponding author}
\ead{fruhling@bc.edu}
\author[mainaddress]{Xiaohan~Yao}
\author[mainaddress]{Alenna~Streeter}
\author[mainaddress]{Fazel~Tafti}
\address[mainaddress]{Department of Physics, Boston College, Chestnut Hill, MA 02467, USA}

\begin{abstract}
Supplemental material includes the analysis of energy dispersive x-ray spectroscopy (EDX) on all samples used in this study and complementary analysis of the magnetocaloric effect (MCE) at low temperatures and in high-entropy alloys containing 4 rare-earths and 5 rare-earths. 
\end{abstract}
\begin{keyword}
Magnetocaloric effect\sep Kagome lattice\sep Rare-earth 
\end{keyword}
\end{frontmatter}

\newpage
\section{EDX analysis}
\begin{figure}
  \includegraphics[width=0.8\textwidth]{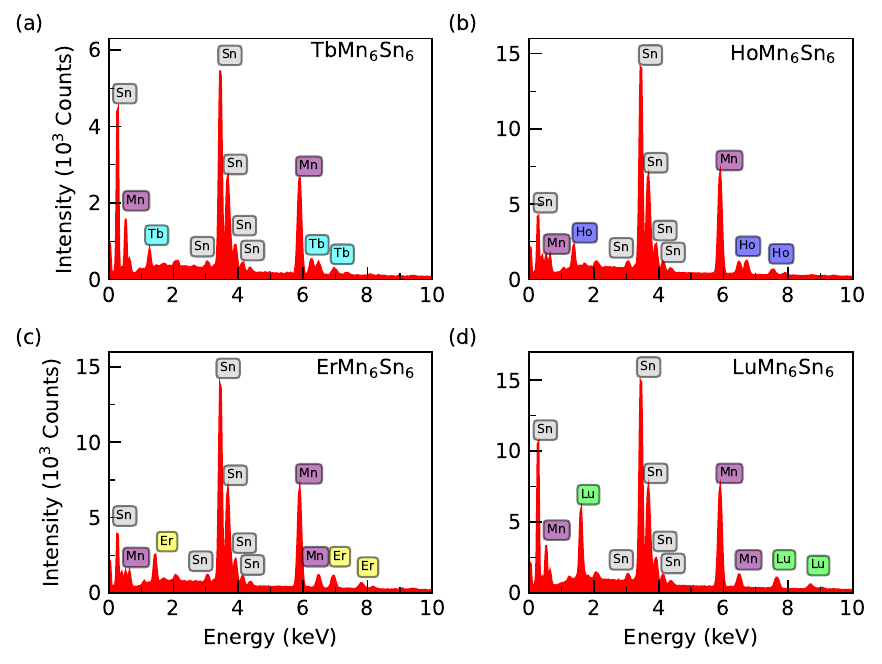}
  \caption{\label{fig:EDX}
   Representative EDX spectra  of (a) TbMn$_6$Sn$_6$, (b) HoMn$_6$Sn$_6$, (c) ErMn$_6$Sn$_6$, and (d) LuMn$_6$Sn$_6$. 
   The compositional analysis in Table~\ref{tab:EDX} comes from averaging the results of a few spectra on each sample.
  }
\end{figure}

The chemical composition of $R$Mn$_6$Sn$_6$ samples in this study, including the high-entropy alloys, were measured by energy dispersive x-ray spectroscopy (EDX) using a scanning electron microscope (SEM).
The electron beam had a current of 0.4~nA and was accelerated under a voltage of 20~keV.
Three different areas were scanned on each sample to calculate the average and standard deviation of the atomic percentage in each sample.
The final compositions with errors are reported in Table.~\ref{tab:EDX}.
Representative spectra from each $R$Mn$_6$Sn$_6$ material are presented in Fig.~\ref{fig:EDX}.
The EDX spectrum of HEA-6 was presented in the main text (Fig.~5a). The spectra of the two high-entropy alloys with mixes of fewer rare-earths are presented in Figs.~\ref{fig:HEA-4}a and ~\ref{fig:HEA-5}a.

\begin{table}[htbp]
\begin{adjustwidth}{-2.5cm}{}
  \centering
    \caption{\label{tab:EDX}EDX results tabulated for 7 compositions used in this study.}
    \begin{adjustbox}{max width=6.5in}
      \begin{tabular}{l|l}
        \hline
        \hline
        Nominal composition & EDX result \\
        \hline
        TbMn$_{6}$Sn$_{6}$    & TbMn$_{6.6(1)}$Sn$_{6.4(1)}$   \\
        HoMn$_{6}$Sn$_{6}$    & HoMn$_{6.4(2)}$Sn$_{6.2(1)}$   \\
        ErMn$_{6}$Sn$_{6}$    & ErMn$_{6.07(4)}$Sn$_{5.83(5)}$ \\
        LuMn$_{6}$Sn$_{6}$    & LuMn$_{6.1(1)}$Sn$_{6.0(2)}$   \\
        (Gd$_{0.25}$Tb$_{0.25}$Dy$_{0.25}$Ho$_{0.25}$)Mn$_{6}$Sn$_{6}$    & (Gd$_{0.29(4)}$Tb$_{0.25(3)}$Dy$_{0.21(4)}$Ho$_{0.25(4)}$)Mn$_{6.7(2)}$Sn$_{6.5(3)}$  \\
        (Gd$_{0.2}$Tb$_{0.2}$Dy$_{0.2}$Ho$_{0.2}$Er$_{0.2}$)Mn$_{6}$Sn$_{6}$   & (Gd$_{0.22(3)}$Tb$_{0.20(4)}$Dy$_{0.18(3)}$Ho$_{0.20(3)}$Er$_{0.20(1)}$)Mn$_{6.8(3)}$Sn$_{6.4(3)}$  \\
        (Gd$_{0.17}$Tb$_{0.17}$Dy$_{0.17}$Ho$_{0.17}$Er$_{0.17}$Lu$_{.17}$)Mn$_{6}$Sn$_{6}$   & (Gd$_{0.21(2)}$Tb$_{0.17(2)}$Dy$_{0.15(2)}$Ho$_{0.16(2)}$Er$_{0.19(1)}$Lu$_{0.13(2)}$)Mn$_{6.4(3)}$Sn$_{6.2(3)}$  \\
        \hline
        \hline
      \end{tabular}
    \end{adjustbox}
    \end{adjustwidth}
\end{table}

\section{Magnetocaloric effect at low temperatures}

\begin{figure*}
  \includegraphics[width=\textwidth]{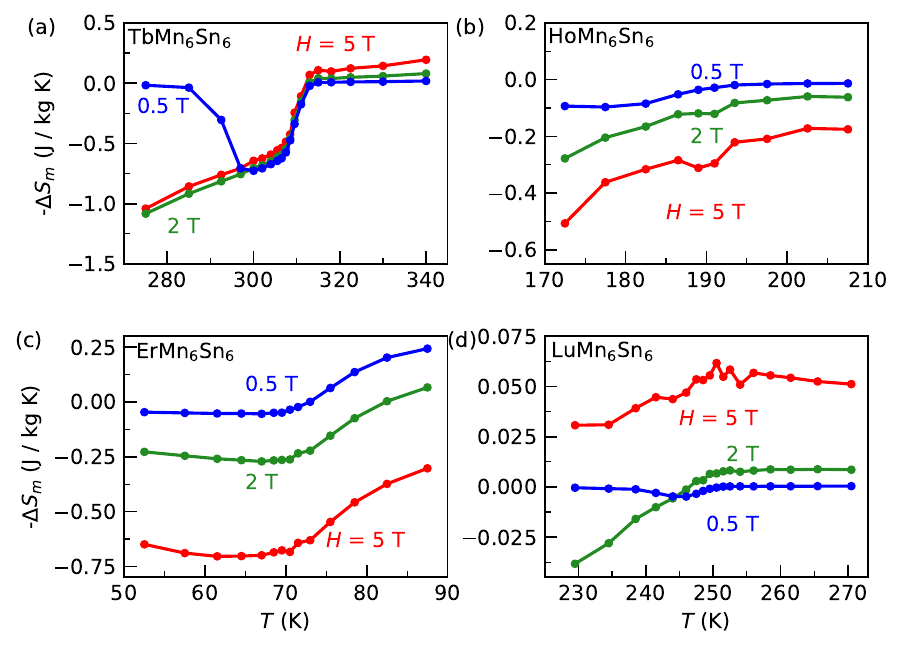}
  \caption{\label{fig:MCE_LT}
  Magnetocaloric effect evaluated around the low temperature transition in (a) TbMn$_6$Sn$_6$ (b) HoMn$_6$Sn$_6$ (c) ErMn$_6$Sn$_6$ and (d) LuMn$_6$Sn$_6$. The red, green, and blue curves correspond to the magnetic entropy integrated with Hmax=5, 2, and 0.5 T, respectively. All measurements are done with field in the ab-plane.
  }
\end{figure*}

The MCE was characterized in the vicinity of the low temperature transitions. In Fig.~\ref{fig:MCE_LT} the red, blue, and green curves trace the change in entropy up to $H_{max}$=5, 2, and 0.5 T respectively. At these lower temperatures the compounds show a reduced MCE and in the case of TbMn$_6$Sn$_6$, HoMn$_6$Sn$_6$, and ErMn$_6$Sn$_6$ the $-\Delta S_m$ with the largest magnitude is negative. Negative values of $-\Delta S_m$ are not useful for applications as they would heat the thermal load instead of cooling it. In Fig.~\ref{fig:MCE_LT}d LuMn$_6$Sn$_6$ shows a positive $-\Delta S_m$ at 5 T, however the  $-\Delta S_m$ values are a full order of magnitude smaller than at the high temperature transition of LuMn$_6$Sn$_6$ making the low temperature transition irrelevant for applications.

\section{Magnetocaloric effect of High Entropy Alloys}

\begin{figure*}
\centering
  \includegraphics[width=0.8\textwidth]{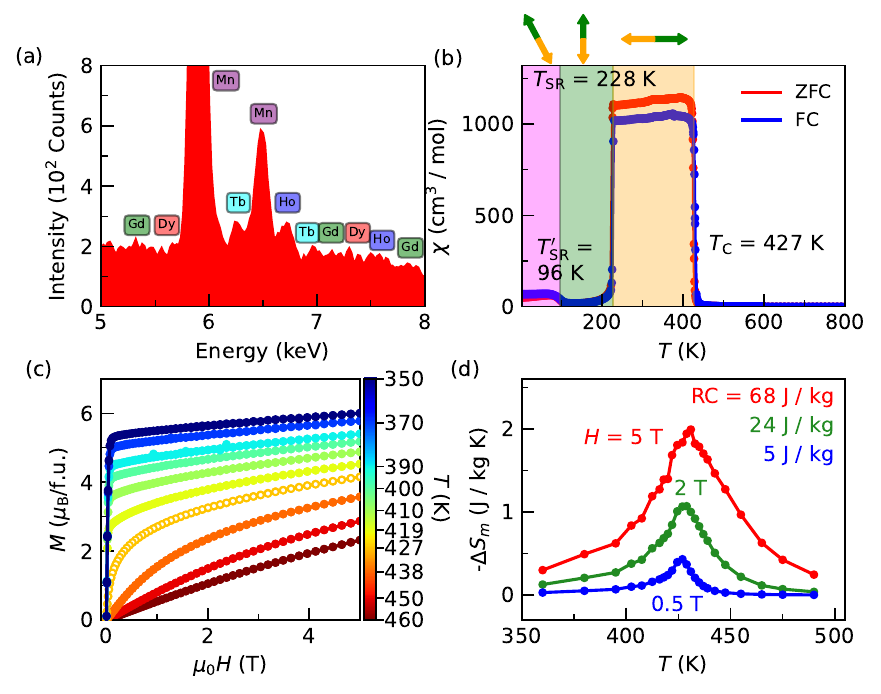}
  \caption{\label{fig:HEA-4}
   (a) EDX spectrum confirms the presence of four rare-earth elements in the 4R-HEA. 
   (b) Susceptibility versus temperature shows an FIM transition at high temperatures and two more transitions (spin reorientations) at lower temperatures. 
   (c,d) The Magnetization curves and MCE behavior of 4R-HEA are comparable to those of pure \ch{$R$Mn6Sn6} systems with FIM ordering such as \ch{TbMn6Sn6} and \ch{HoMn6Sn6}.
  }
\end{figure*}

\begin{figure}
\centering
  \includegraphics[width=0.8\textwidth]{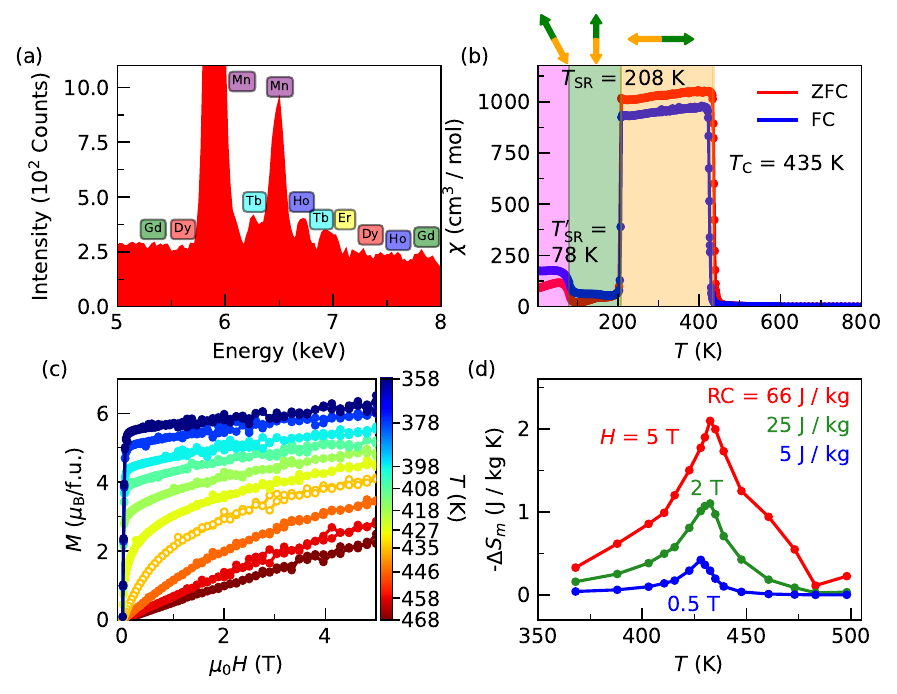}
  \caption{\label{fig:HEA-5}
   (a) EDX spectrum confirms the presence of five rare-earth elements in the 5R-HEA. 
   (b) Susceptibility versus temperature shows an FIM transition at high temperatures and two more transitions (spin reorientations) at low temperatures. 
   (c,d) The Magnetization curves and MCE behavior of 5R-HEA are comparable to those of the 4R-HEA and pure \ch{$R$Mn6Sn6} systems with FIM ordering.
  }
\end{figure}

The MCE of two high entropy alloys with fewer rare-earths included are reported here. We refer to (Gd$_{0.25}$Tb$_{0.25}$Dy$_{0.25}$Ho$_{0.25}$)Mn$_{6}$Sn$_{6}$ as HEA-4 and (Gd$_{0.2}$Tb$_{0.2}$Dy$_{0.2}$Ho$_{0.2}$Er$_{0.2}$)Mn$_{6}$Sn$_{6}$ as HEA-5. Fig.~\ref{fig:HEA-4} and Fig.~\ref{fig:HEA-5} show that these alloys behave similar to HEA-6 with a high temperature FIM ordering with two lower temperature spin reorientation transitions. The MCE in HEA-4 and HEA-5 are almost equivalent in terms of both $-\Delta S_{\text{MAX}}$ and $RC$. The $RC$ in these alloys is reduced, however, from that of HEA-6, indicating that there may be some improvement with a higher mixture of metals.

